\documentclass[12pt]{elsart}

\usepackage{graphicx}
\usepackage{subfigure}
\usepackage{endfloat}
\usepackage{amsmath}
\usepackage{amssymb}
\usepackage{cite}

\journal{Physica A}

\begin{document}

\begin{frontmatter}

\title{Phase diagram of a mixed spin-1 and spin-3/2 Ising ferrimagnet}
\author{M. \v{Z}ukovi\v{c}\corauthref{cor}},
\ead{milan.zukovic@upjs.sk}
\author{A. Bob\'ak}
\ead{andrej.bobak@upjs.sk}
\address{Department of Theoretical Physics and Astrophysics, Faculty of Science, \\ P. J. \v{S}af\'arik University, Park Angelinum 9, 041 54 Ko\v{s}ice, Slovak Republic}
\corauth[cor]{Corresponding author.}

\begin{abstract}
\hspace*{5mm} Critical and compensation properties of a mixed spin-$1$ and spin-$3/2$ Ising ferrimagnet on a square lattice are investigated by standard and histogram Monte Carlo simulations. The critical temperature is studied as a function of a single-ion anisotropy strength. The second order of the phase transition is established by finite-size scaling for the entire boundary. Some previously obtained results, such as a tricritical point, predicted by the mean field theory (MFT) and effective field theory (EFT), or a first-order transition line separating two different ordered phases, obtained by the cluster variational theory (CVT), are deemed artifacts of the respective approximations. So is a reentrant phenomenon produced by CVT. Nevertheless, the multicompensation behavior predicted by MFT and EFT was confirmed.
\end{abstract}


\end{frontmatter}

\section{Introduction}
\hspace*{5mm} There have been a number of theoretical studies of mixed-spin Ising ferrimagnets as simple models of certain types of molecular-based magnetic materials \cite{iwas84,vero88,kahn93,kane98a,kane98b}. Besides other interesting properties, such as appearance of a multicritical behavior, they attract attention due to so-called compensation behavior with possible technological applications. Such a phenomenon occurs at a compensation temperature, i.e., the temperature below the critical point at which the sublattice magnetizations completely cancel out and the total magnetization changes sign. Generally speaking, from the previous studies it can be concluded that the higher values of spins and the more complex lattice topologies used the richer behavior can be expected. However, the increasing complexity usually requires increasing simplifications in the approaches for the problem to be manageable. Thus nonperturbative approaches, such as exact \cite{gonc85,lipo95,jasc98,dakh98,jasc05} or Monte Carlo (MC) \cite{zhan93,buen97,godo04,selk00}, are so far limited to the simplest cases of either the smallest spin values (i.e., mixed spin-1/2 and spin-1) or the lattice topology (e.g., honeycomb or Bethe lattices). As a consequence, the behavior of the mixed spin-1/2 and spin-1 Ising system is rather well understood, while there are still disagreements among different theoretical investigations of the mixed-spin systems with higher spin values, including the spin-1 and spin-3/2 case. The disagreements arise from the fact that due to higher complexity mostly different approximative schemes with various degrees of approximation have been employed, such as mean field theory (MFT) \cite{abub01}, effective field theory (EFT) \cite{boba98,boba00,boba02} and cluster variational theory (CVT) \cite{tuck01}. These approximative approaches have been previously shown to produce some artifacts, such as a tricritical point, predicted by MFT \cite{kane91} and EFT \cite{boba97} for the mixed spin-1/2 and spin-1 system on a square lattice, that were not reproduced either in numerical transfer matrix \cite{buen97} or MC studies \cite{zhan93,buen97,selk00}. To our knowledge, the results of these approximative studies for the mixed spin-1 and spin-3/2 system have so far been verified by MC simulations only for the case of a simple cubic lattice \cite{naka02}. However, for this particular case, there were basically no qualitative disagreements among the conclusions drawn from the respective studies and the MC results only confirmed the previously obtained results. On the other hand, for the cases of honeycomb and square lattices, there are qualitative differences in the results, which have not been resolved yet. \\
\hspace*{5mm} The objective of this study is to focus on the case of the mixed spin-1 and spin-3/2 Ising system with a uniform single-ion anisotropy on a square lattice, for which the differences between the MFT, EFT and CVT results are the most prominent. In particular, we aim to answer the following questions: (1) Are the phase transitions of second order for the entire range of the anisotropy strength, as predicted by CVT, or is there a tricritical point separating lines of the second- and first-order transitions, as predicted by both MFT and EFT? (2) Is there a reentrant phenomenon in the second-order phase boundary, as suggested by the CVT results but not by MFT nor EFT? (3) Does the line of first-order transitions situated within the ordered ferrimagnetic region, obtained by CVT but not by MFT nor EFT, really exist? (4) Can the system display up to two compensation points, as predicted by both MFT and EFT (not investigated by CVT)?

\section{Model and Monte Carlo simulations}
\hspace*{5mm} The model of the mixed spin-1 and spin-3/2 Ising system on the square lattice is described by the Hamiltonian 
\begin{eqnarray}
\label{Hamil}
H = -J\sum_{(i,j)}S_i^AS_j^B - D_A\sum_i(S_i^A)^2 - D_B\sum_j(S_j^B)^2,
\end{eqnarray}
where $S_i^A = \pm \frac{3}{2},$ $\pm \frac{1}{2}$ for $A$ ions, $S_j^B = \pm 1, 0$ for $B$ ions, $J < 0$ is the nearest-neighbor coupling parameter between the ions on A and B sublattices, and $D_A$, $D_B$ are the single-ion anisotropies acting on the spin-3/2 and spin-1 ions, respectively. In this study we will consider the anisotropy of a uniform strength i.e., $D\equiv D_A=D_B$. \\
\hspace*{5mm} A simulated $L \times L$ square lattice consists of two interpenetrating sublattices, each one comprising $L^2/2$ sites. We consider linear lattice sizes ranging from $L=20$ up to $L=200$ with the periodic boundary conditions imposed. Initial spin states are randomly assigned and the updating follows the Metropolis dynamics. The lattice structure and the short range of the interactions enable vectorization of the algorithm. Since the spins on one sublattice interact only with the spins on the other, each sublattice can be updated simultaneously. Thus one sweep through the entire lattice involves just two sublattice updating steps. For thermal averaging, we typically consider $N = 10^5$ MC sweeps in the standard and up to $N = 10^7$ MC sweeps in the histogram MC simulations \cite{ferr88,ferr89}, after discarding another $20$\% of these numbers for thermalization. To assess uncertainity, we perform $10$ runs, using different random initial configurations. Then the errors of the calculated quantities are determined from the values obtained for those runs as twice of the standard deviations.\\
\hspace*{5mm} We calculate the internal energy per site $e=E/L^2=\langle H \rangle/L^2$ and the sublattice magnetizations per site
\begin{equation}
\label{Magn_A}
m_A = 2\langle M_A \rangle/L^2 = 2\Bigg\langle \Bigg|\sum_{A}S_{i}^{A}\Bigg| \Bigg\rangle/L^2,
\end{equation}
\begin{equation}
\label{Magn_B}
m_B = 2\langle M_B \rangle/L^2 = 2\Bigg\langle -\Bigg|\sum_{B}S_{j}^{B}\Bigg| \Bigg\rangle/L^2,
\end{equation}
\noindent where $\langle\cdots\rangle$ denotes thermal averages. The total magnetization per site is defined as
\begin{equation}
\label{Magn_dir}
m = \langle M \rangle/L^2 = \langle M_A + M_B \rangle/L^2.
\end{equation}
Since for ferrimagnets $m$ can vanish within the ordered phase at the compensation temperature, as an order parameter it is useful to define the staggered magnetization per site as
\begin{equation}
\label{Magn_stag}
m_s = \langle M_s \rangle/L^2 = \langle M_A - M_B \rangle/L^2.
\end{equation}
Further, the following quantities which are functions of the parameters $E$ or/and $O$ ($=\ M,\ M_s$) are defined: 
the specific heat per site $c$
\begin{equation}
\label{eq.c}c=\frac{\langle E^{2} \rangle - \langle E \rangle^{2}}{L^2k_{B}T^{2}}\ ,
\end{equation}
the direct $(O=M)$ and staggered $(O=M_s)$ susceptibilities per site
$\chi_{O}$
\begin{equation}
\label{eq.chi}\chi_{O} = \frac{\langle O^{2} \rangle - \langle O \rangle^{2}}{L^2k_{B}T}\ ,
\end{equation}
the logarithmic derivatives of $\langle O \rangle$ and $\langle O^{2} \rangle$ with respect to
$\beta=1/k_{B}T$
\begin{equation}
\label{eq.D1}D_{1O} = \frac{\partial}{\partial \beta}\ln\langle O \rangle = \frac{\langle OE
\rangle}{\langle O \rangle}- \langle E \rangle\ ,
\end{equation}
\begin{equation}
\label{eq.D2}D_{2O} = \frac{\partial}{\partial \beta}\ln\langle O^{2} \rangle = \frac{\langle O^{2} E
\rangle}{\langle O^{2} \rangle}- \langle E \rangle\ ,
\end{equation}
the Binder parameter $U$
\begin{equation}
\label{eq.U}U = 1-\frac{\langle M^{4}\rangle}{3\langle M^{2}\rangle^{2}} \ ,
\end{equation}
and the fourth-order energy cumulant $V$
\begin{equation}
\label{eq.V}V = 1-\frac{\langle E^{4}\rangle}{3\langle E^{2}\rangle^{2}}\ .
\end{equation}
\hspace*{5mm} The above quantities are useful for localization of a phase transition as well as for determination of its
nature. For example, the transition temperatures can be estimated from the locations of the peaks of the response functions $c$ and $\chi_O$ or, with a higher precision, from the intersection of the Binder parameter $U$ curves for different lattice sizes $L$. The first-order character of the transition can be judged from the presence of discontinuities in the magnetizations and energy, as well as from hysteresis loops. Furthermore, the order of the transition can be reliably established by a finite-size scaling (FSS) analysis. For instance, the energy cumulant $V$ exhibits a minimum near critical temperature $T_{c}$, which achieves the
value $V^{*}=\frac{2}{3}$ in the limit $L\rightarrow \infty$ for a second-order transition, while $V^{*}<\frac{2}{3}$ is expected for
a first-order transition \cite{ferr88,ferr89}. Temperature-dependences of a variety of thermodynamic quantities
display extrema at the $L$-dependent transition temperatures, which at a second-order transition are known to scale with a
lattice size as, for example:
\begin{equation}
\label{eq.scalchi}\chi_{O,max}(L) \propto L^{\gamma_{O}/\nu_{O}}\ ,
\end{equation}
\begin{equation}
\label{eq.scalD1}D_{1O,max}(L) \propto L^{1/\nu_{O}}\ ,
\end{equation}
\begin{equation}
\label{eq.scalD2}D_{2O,max}(L) \propto L^{1/\nu_{O}}\ ,
\end{equation}
\noindent where $\nu_{O}$ and $\gamma_{O}$ represent the critical exponents of the correlation length and susceptibility, respectively. More precise locations of the extrema used in FSS can be obtained from histogram MC simulations \cite{ferr88,ferr89}, performed at criticality for each lattice size. In the case of a first-order transition, typically a bimodality appears in the $P(E)$ and $P(O)$ histograms as $L$ increases and the quantities (\ref{eq.c})$-$(\ref{eq.V}) display a volume-dependent scaling, $\propto L^{2}$.

\section{Results and discussion}
\begin{figure}
\centering
\includegraphics[scale=0.7]{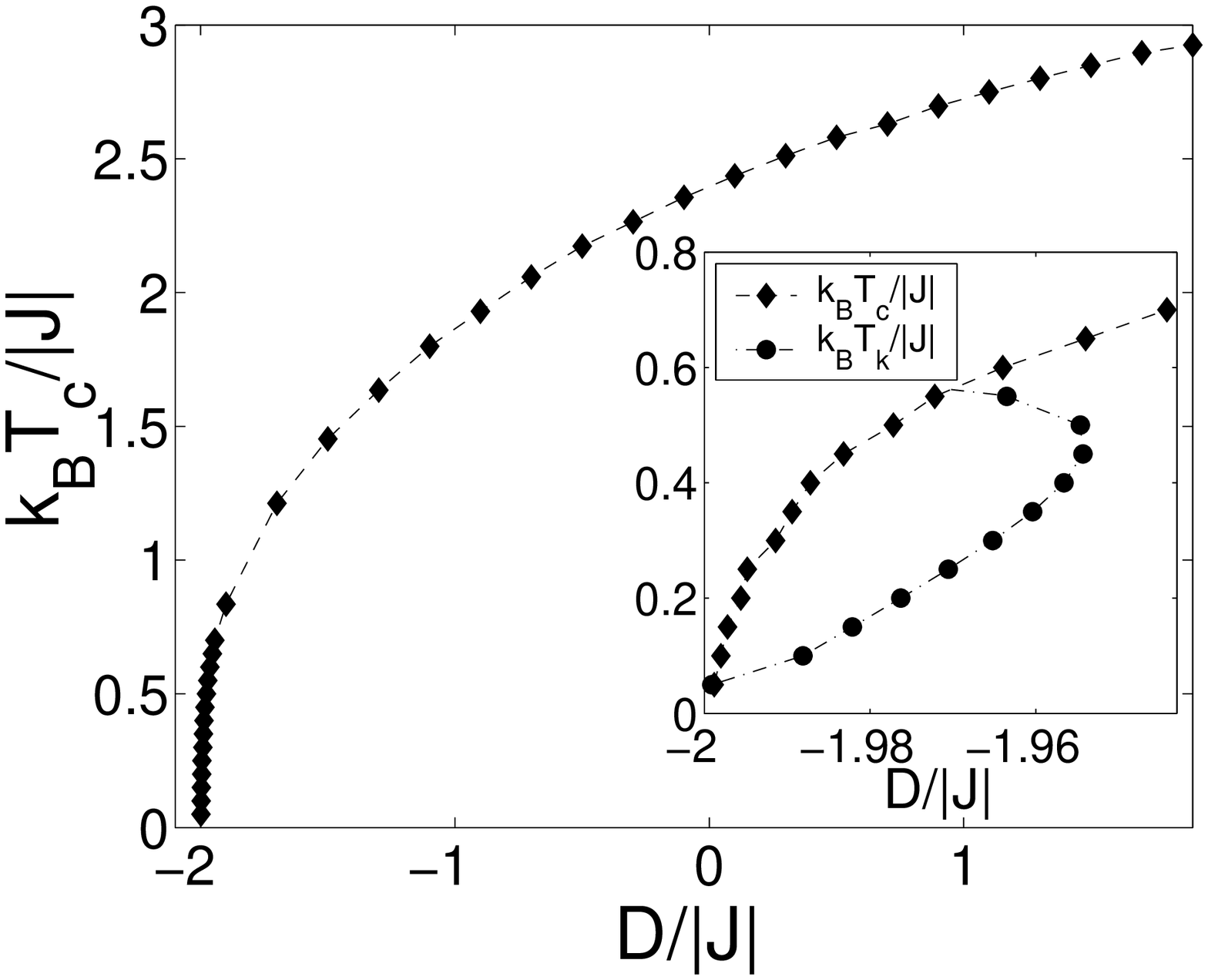}%
\caption{The critical temperature $k_BT_c/|J|$ as a function of the single-ion anisotropy strength $D/|J|$. The inset shows the enlarged picture in a low-temperature region with both the critical ($k_BT_c/|J|$) and compensation ($k_BT_k/|J|$) temperature curves.}
\label{fig:D-Tc_comb}
\end{figure}

\hspace*{5mm} The critical temperature as a function of the anisotropy strength $D$, shown in Fig.~\ref{fig:D-Tc_comb}, was determined from the positions of the staggered susceptibility $\chi_{M_s}$ peaks obtained by standard MC simulations for a fixed lattice size of $L=40$. We checked that the peaks positions (the temperatures at which they occur) do not significantly change when larger $L$ is used. For the case of $D_A=D_B=0$ we estimated the critical temperature with relatively high precision from the Binder parameter analysis, using the lattice sizes from $L=40$ up to $200$ and $N = 5\times10^{6}$ MC sweeps, as $k_BT_c/|J|=2.362 \pm 0.003$. In line with the MFT \cite{abub01}, EFT \cite{boba98,boba00,boba02} and CVT \cite{tuck01} results, the critical temperature decreases as the anisotropy is decreased from positive to negative values and eventually vanishes at the exact value of $D/|J|=-2$. However, in contrast to the CVT results, no signs of the reentrant phenomenon was observed near $D/|J|=-2$. Namely, as temperature is increased from low values there is a single order-disorder phase transition for each value of $D/|J|>-2$ and no phase transition below $D/|J|=-2$.

\begin{figure}
\centering
\includegraphics[scale=0.7]{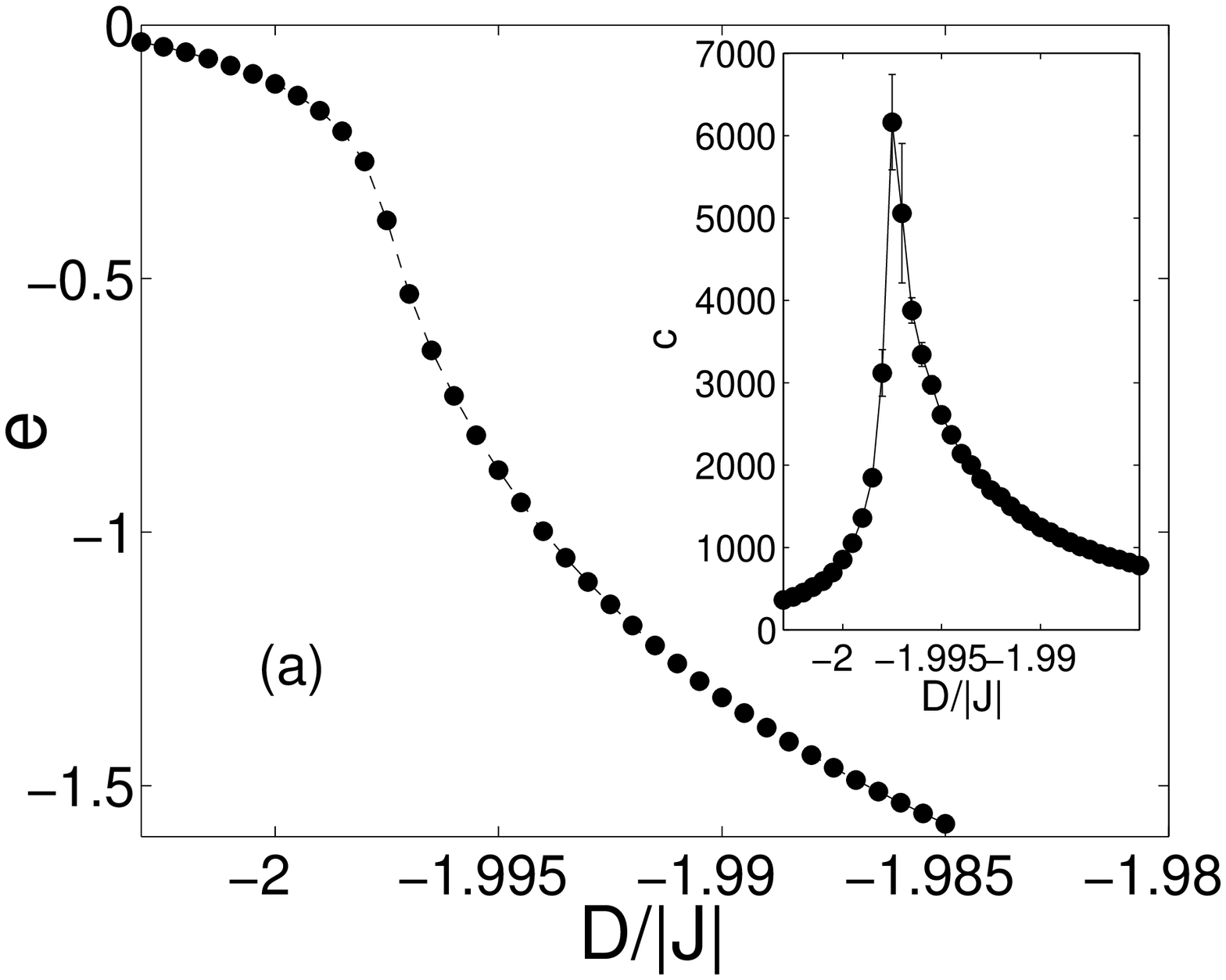}
\includegraphics[scale=0.7]{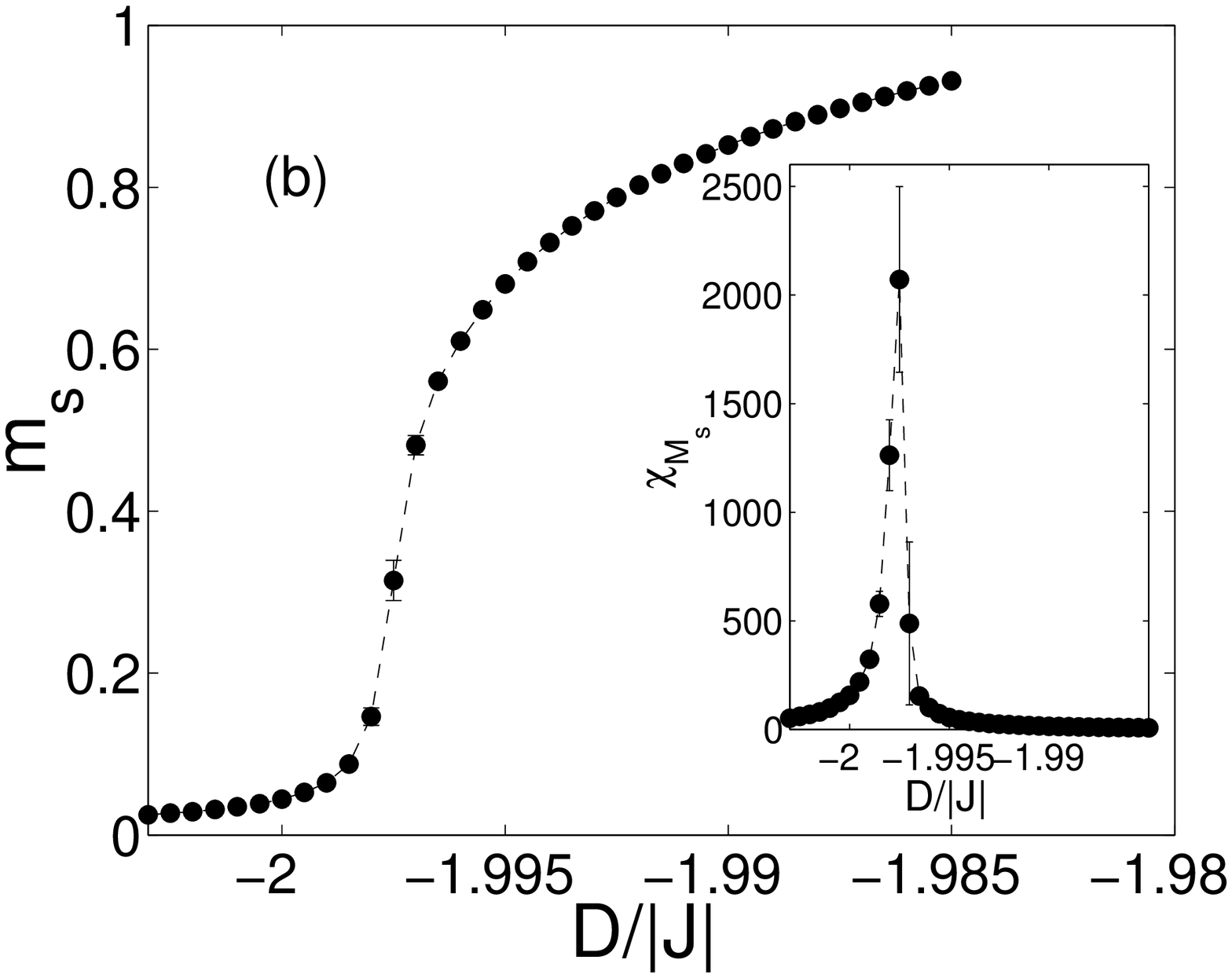}
\caption{(a) The internal energy $e$ and specific heat $c$ (inset) per site and (b) the staggered magnetization $m_s$ and staggered susceptibility $\chi_{M_s}$ (inset) per site, as functions of the anisotropy strength $D/|J|$, at $k_BT/|J|=0.1$.}
\label{fig:ems-D}
\end{figure}

\begin{figure}
\centering
\includegraphics[scale=0.7]{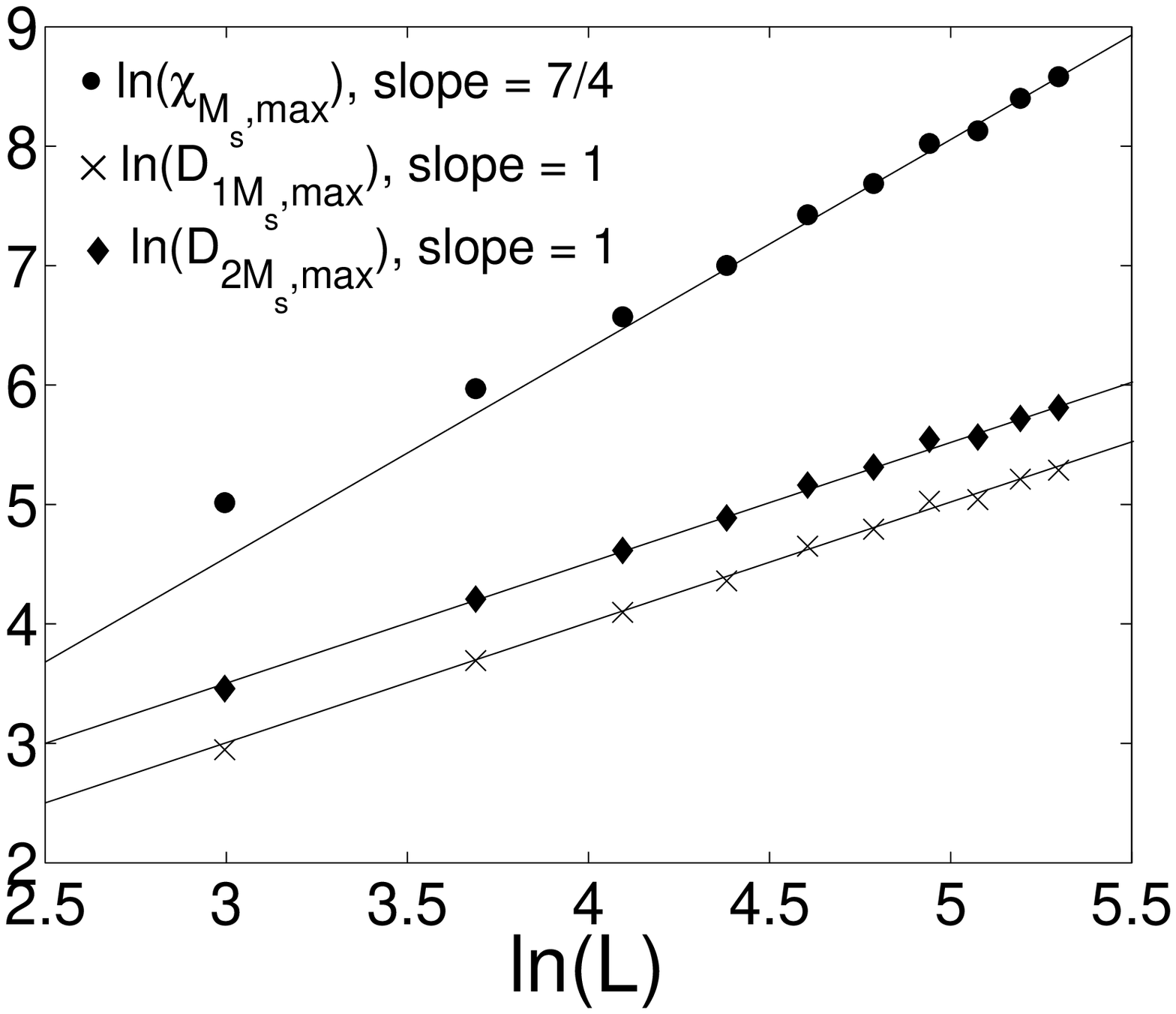}%
\caption{Scaling behavior of the maxima of the quantities (\ref{eq.chi})$-$(\ref{eq.D2}) for the parameter $O=M_s$ i.e., $\chi_{M_s,max}$, $D_{1M_s,max}$ and $D_{2M_s,max}$, in a log-log plot at the critical point $(D_c/|J|,k_BT_c/|J|) = (-1.9975,0.1)$. The slopes represent the values of the standard 2D Ising exponents ratios $\gamma_{I}/\nu_{I}=7/4$ and $1/\nu_{I}=1$.}
\label{fig:LogLog}
\end{figure}

\begin{figure}
\centering
\includegraphics[scale=0.7]{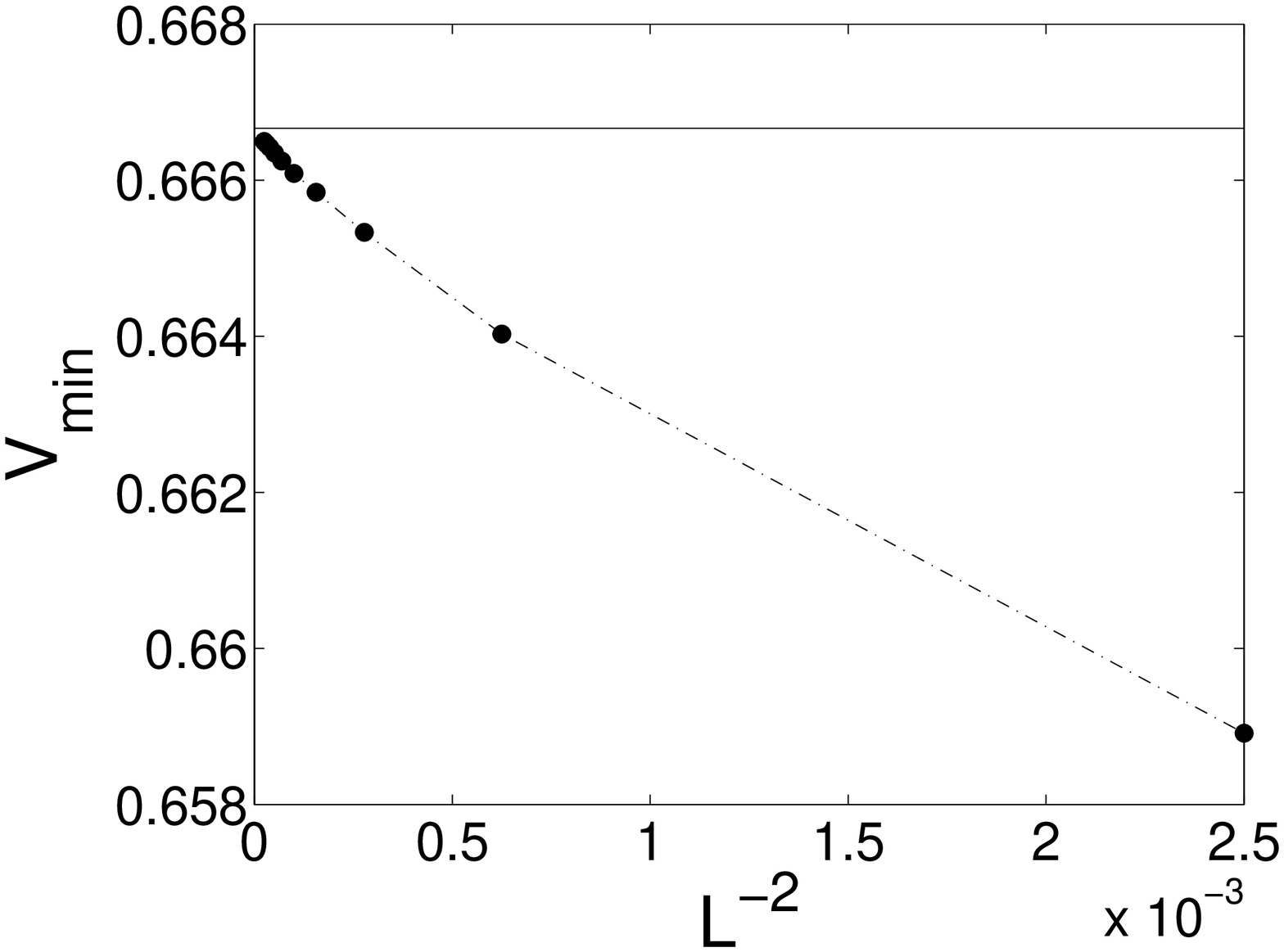}%
\caption{The minimum of the fourth-order energy cumulant $V$ as a function of $L^{-2}$. The solid horizontal line denotes $V^{*}=2/3$.}
\label{fig:V-L}
\end{figure}

\hspace*{5mm} Furthermore, no indications of the additional first-order transition line within the ordered phase at low temperatures, found in the CVT study, were observed. In Fig.~\ref{fig:ems-D} we show anisotropy dependences of the staggered magnetization $m_s$ and energy $e$, along with the respective response functions, $\chi_{M_s}$ and $c$, shown in insets, at $k_BT/|J|=0.1$. It is well known that discontinuities in physical quantities associated with a first-order transition tend to smear out in MC simulations for smaller lattice sizes. Therefore, in order to detect first-order transition features, we chose a relatively large size of $L=120$ and also checked for possible loops in the curves obtained by increasing and decreasing $D$. According to \cite{tuck01}, we should observe two anomalies associated with two phase transitions: the first-order one between two ordered phases and the second-order one between the ferrimagnetic and paramagnetic phases. However, as evidenced in Fig.~\ref{fig:ems-D}, only one anomaly can be observed and, therefore, we conclude that there is only one phase transition which is from the ordered ferrimagnetic to the paramagnetic phase. This scenario agrees with the MFT and EFT predictions. However, the latter insist that in this low-temperature region, i.e., below the tricritical point $(D_t/|J|,k_BT_t/|J|)=(-1.9730,1.1606)$ from MFT and $(D_t/|J|,k_BT_t/|J|)=(-1.9981,0.6175)$ from EFT, the transition is first-order, which does not appear so in our simulations, since no discontinuities nor loops in the staggered magnetization and internal energy curves can be seen.\\ \hspace*{5mm} Nevertheless, the respective spike-like response functions, which are somewhat reminiscent of a first-order transition, prompted us to further explore the order of the transition by running more extensive histogram MC simulations for even larger sizes $L$ and performing FSS analysis. We considered the lattice sizes up to $L=200$ and the number of MC sweeps $N=10^7$, however, we did not detect any signs of a bimodal distribution in the energy or staggered magnetization histograms at criticality that would signal a first-order transition. On the contrary, FSS of the quantities (\ref{eq.chi})$-$(\ref{eq.D2}), shown in Fig.~\ref{fig:LogLog}, indicates that the calculated susceptibility and correlation length critical exponents, $\gamma_{M_s}$ and $\nu_{M_s}$, respectively, agree well with the standard 2D Ising ones, i.e., $\gamma_I=7/4$ and $\nu_I=1$. In the susceptibility exponent dependence we can see that a linear regime in the log-log behavior is established only at $L\approx 80$ and, therefore, sufficiently large lattice sizes should be used. The second-order nature of the transition is also confirmed by the scaling of the minima of the fourth-order energy cumulant $V$ (Fig.~\ref{fig:V-L}), which tend to the value $V^*=2/3$ for $L \rightarrow \infty$ and apparently do not scale with volume, as expected for the second-order transition. Such thorough investigations were not done below $k_BT/|J|=0.1$ and we cannot rule out the possibility of the tricritical point existence at still lower temperatures but we deem it unlikely and presume that the transition remains second-order at all temperatures. 

\begin{figure}
\centering
\includegraphics[scale=0.7]{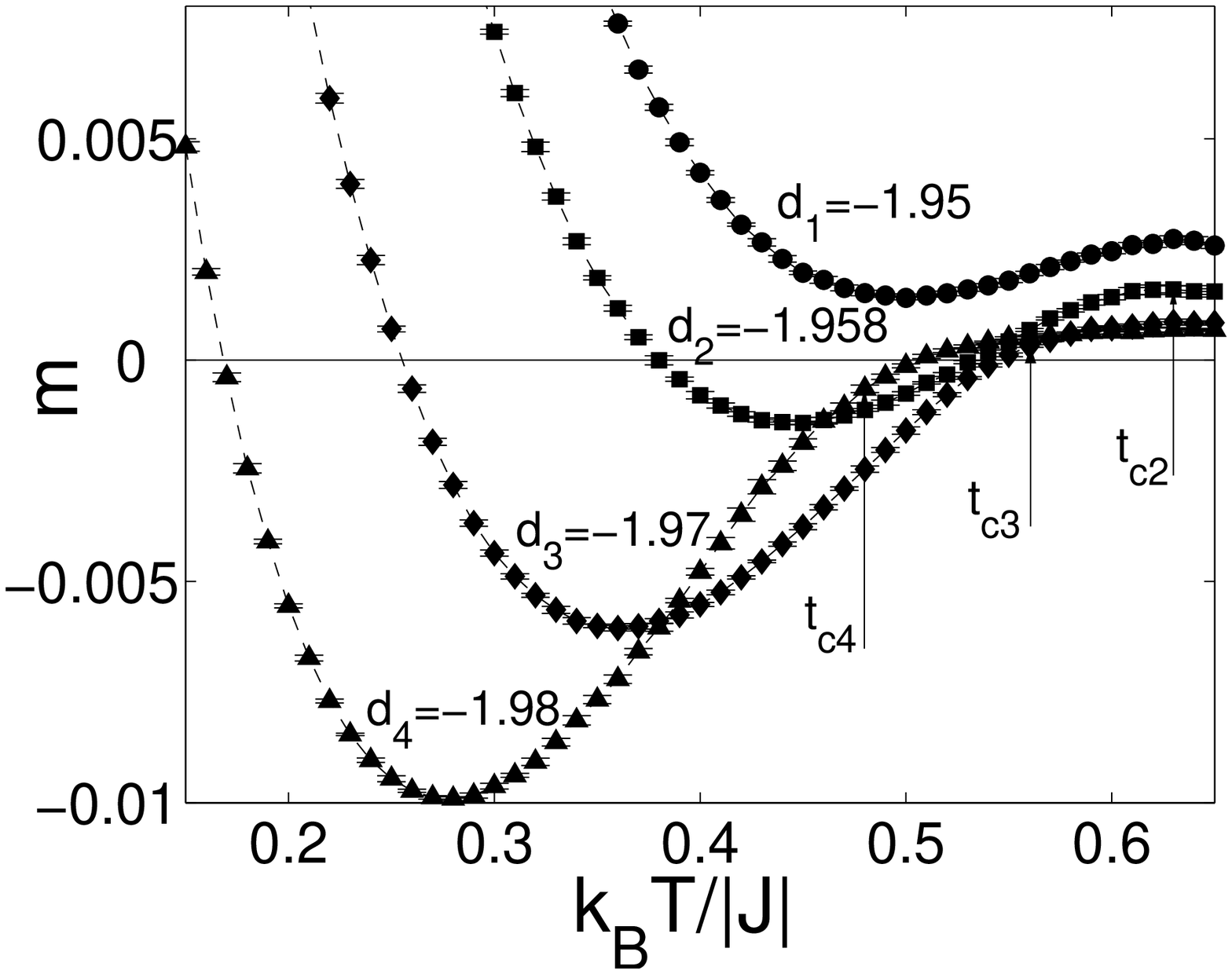}%
\caption{The total magnetization vs temperature curves for several values of the anisotropy strength $d \equiv D/|J|$. There is no compensation point for $d_1=-1.95$, two compensation points for $d_2=-1.958$ and $d_3=-1.97$ and one compensation point for $d_4=-1.98$. $t_{ci}\equiv k_BT_{ci}/|J|$, $i=1,2,3,4$, denote the critical temperatures corresponding to the respective values of $d_i$.}
\label{fig:m-T_reent}
\end{figure}

\hspace*{5mm} Finally, we verified the MFT \cite{abub01} and EFT \cite{boba02} predictions about the multicompensation behavior of the system. Namely, these theories predict the existence of two compensation points at the anisotropy values close to $D/|J|=-1.96$. Such a behavior was indeed confirmed in our MC simulations. In Fig.~\ref{fig:m-T_reent} we plot the total magnetization vs temperature curves for several values of the anisotropy strength $d \equiv D/|J|$. While for $d_1=-1.95$ there is no compensation point, for the values of $d_2=-1.958$ and $d_3=-1.97$, there are two compensation points below the respective critical temperatures $t_{c2} (\equiv k_BT_{c2}/|J|)$ and $t_{c3}$. For $d_4=-1.98$, there is only one compensation point below $t_{c4}$. The entire curve of the compensation temperature $k_BT_k/|J|$ as a function of the anisotropy strength $D/|J|$ is shown in the inset of Fig.~\ref{fig:D-Tc_comb}.

\section{Conclusions}
\hspace*{5mm} We have studied the critical and compensation properties of the mixed spin-$1$ and spin-$3/2$ Ising ferrimagnet with a uniform single-ion anisotropy on a square lattice by Monte Carlo simulations. We focused on several contradictory results previously obtained by different approximative approaches. In particular, we checked if the system for sufficiently large anisotropy strength can display a reentrant phenomenon and additional first-order phase transition within the ordered ferrimagnetic phase, as predicted by CVT. Neither of these predictions was confirmed. Instead, we found just one order-disorder phase boundary as a single-valued function of the anisotropy within the entire range of values, in agreement with the MFT and EFT results. Thus, while for the system on a cubic lattice the qualitative features predicted by CVT were confirmed by the MC simulations \cite{naka02}, they were not confirmed on a square lattice. On the other hand, our findings neither support the MFT and EFT predictions about the existence of a tricritical point at which the transition would change to the first order one. Nevertheless, the multicompensation behavior with two compensation points observed within the MFT and EFT studies was reliably verified in the current MC simulations.

\section*{Acknowledgments}
This work was supported by the Scientific Grant Agency of Ministry of Education of Slovak Republic (Grant VEGA No. 1/0128/08).


\end{document}